\documentclass[prl,twocolumn,floatfix,superscriptaddress]{revtex4-1}
\usepackage{dcolumn,amsmath}
\usepackage{graphicx}
\usepackage{bm}
\usepackage{hyperref}

\usepackage[utf8]{inputenc}
\usepackage[T1]{fontenc}

\begin{document}

\title{Nuclear magnetization distribution effect in molecules: Ra$^+$ and RaF hyperfine structure}

\date{August 4, 2020}

\begin{abstract}
Recently the first laser spectroscopy measurement of the radioactive RaF molecule has been reported by Ruiz \textit{et al.} [Nature \textbf{581}, 396 (2020)]. This and similar molecules are considered to search for the New Physics effects. The radium nucleus is of interest as it is octupole-deformed and has close levels of opposite parity. The preparation of such experiments can be simplified if there are reliable theoretical predictions. It is shown that the accurate prediction of the hyperfine structure of the RaF molecule requires to take into account the finite magnetization distribution inside the radium nucleus. For atoms, this effect is known as the Bohr-Weisskopf (BW) effect. Its magnitude depends on the model of the nuclear magnetization distribution which is usually not well known. We show that it is possible to express the nuclear magnetization distribution contribution to the hyperfine structure constant in terms of one magnetization distribution dependent parameter: BW matrix element for $1s$-state of the corresponding hydrogen-like ion. This parameter can be extracted from the accurate experimental and theoretical electronic structure data for an ion, atom, or molecule without the explicit treatment of any nuclear magnetization distribution model. This approach can be applied to predict the hyperfine structure of atoms and \textit{molecules} and allows one to separate the nuclear and electronic correlation problems. It is employed to calculate the finite nuclear magnetization distribution contribution to the hyperfine structure of the $^{225}$Ra$^+$ cation and $^{225}$RaF molecule. For the ground state of the $^{225}$RaF molecule, this contribution achieves 4\%.
\end{abstract}

\author{Leonid V.\ Skripnikov}
\email{skripnikov\_lv@pnpi.nrcki.ru,\\ leonidos239@gmail.com}
\homepage{http://www.qchem.pnpi.spb.ru}
\affiliation{Petersburg Nuclear Physics Institute named by B.P. Konstantinov of National Research Centre
``Kurchatov Institute'', Gatchina, Leningrad District 188300, Russia}
\affiliation{Saint Petersburg State University, 7/9 Universitetskaya nab., St. Petersburg, 199034 Russia}

\maketitle

\section{Introduction}

Theoretical predictions of hyperfine splittings in heavy atoms and molecules are of great importance for a wide scope of fundamental physical applications~\cite{Safronova:18}. They are used as a test of the accuracy of calculated characteristics that are required for interpretation of experiments to search for effects of violation of spatial parity (P) or spatial parity and time-reversal (T) symmetries of fundamental interactions in atoms~\cite{Safronova:18,Porsev:2009,ginges2017ground,Skripnikov:2020b,GFreview} (see also references therein) and molecules~\cite{Quiney:98,Titov:06amin,Skripnikov:15b,Skripnikov:15a,Sunaga:16,Fleig:17,Borschevsky:2020}  or for the semi-empirical predictions~\cite{Khriplovich:91,Kozlov:95,Kozlov:97c}. The hyperfine splitting in spectra of highly charged ions can be used to test bound-state quantum electrodynamics (QED) in a strong electric and magnetic fields~\cite{beiersdorfer2001hyperfine,Crespo:1998,Shabaev:01a,Volotka:12,Skripnikov:18a,Skripnikov:2020a,Nortershauser:2019}. Accurate theoretical prediction is also required to obtain the values of magnetic moments of short-lived isotopes~\cite{Persson1998,barzakh2012hyperfine,Schmidt:2018,Prosnyak:2020,konovalova2017calculation,atoms6030039}.

In many important cases, the largest uncertainty of the theoretically predicted value of the hyperfine splitting comes from the nuclear part of the problem~\cite{Shabaev:1997,Ginges:2018}. There are two important contributions to the hyperfine splitting due to the finite nucleus size. The first one is caused by the finite change distribution inside the nucleus~\cite{rosenthal1932isotope, crawford1949mf}. It can be calculated with high accuracy as this distribution is known from experiments. The second finite nucleus contribution to the hyperfine splitting is caused by the distribution of magnetization inside the nucleus. In atoms this effect is known as the Bohr-Weisskopf (BW) effect~\cite{bohr1950influence,Senkov:02}. In most cases, it hardly can be calculated with high accuracy due to difficulties of \textit{ab initio} prediction of the magnetization distribution within the nuclear structure theory in case of heavy nuclei as well as the absence of corresponding experimental data.

In Ref.~\onlinecite{Shabaev:01a} it was proposed to use the specific difference of hyperfine splittings in lithium-like and hydrogen-like bismuth to test bound-state QED in strong fields. It was shown that it is possible to choose a linear combination of hyperfine splittings in such a way that the BW contributions cancel. In Ref.~\onlinecite{Ginges:2018} it was proposed to combine experimental and theoretical data on the hyperfine splitting for high lying electronic states of an atom to obtain BW correction for the ground electronic state. The method is valid for nS$_{1/2}$, nP$_{1/2}$ and nP$_{3/2}$  states of alkali-metal atoms and alkali-metal-like ions~\cite{Ginges:2018}. These methods are of great importance for atomic and highly charged ions experiments. For example, the accuracy of the prediction of the hyperfine structure (HFS) constant of $^{133}$Cs is used to estimate the uncertainty of the prediction of the theoretical parameter that is used to interpret the experiment on the $^{133}$Cs~\cite{Wood:1997} atom in terms of fundamental parameters of P-violating forces.

The strongest upper bounds on the electron electric dipole moment (T,P-violating property) has been established in the molecular beam experiment with diatomic ThO molecules~\cite{ACME:18}. Another candidate is the RaF molecule. This molecule can be laser cooled~\cite{Isaev:2010}. This will increase the coherence time and therefore enhance the sensitivity of such an experiment to T,P-violating effects~\cite{Isaev:2010,Isaev:2012,Borschevsky:13,Kudashov:14,Sudip:2016b,Ruiz:2019}. An important feature of this molecule is that it posses the deformed Ra nucleus with close levels of opposite parity~\cite{Butler:2020,Spevak:97,Engel:2003,Cwiok:1991}. Therefore, the T,P-violating effects such as the nuclear Schiff moment are strongly enhanced~\cite{Auerbach:96,Spevak:97,Kudashov:13,Butler:2020,Skripnikov:2020c}. The most elaborated calculation of the nuclear Schiff moment has been performed for the $^{225}$Ra isotope~\cite{Engel:2013}.

Recently, the first laser spectroscopy measurement of the RaF molecule has been performed~\cite{Ruiz:2019}. It is expected that further experimental study of the molecule will allow one to measure magnetic dipole hyperfine constants~\cite{Garcia:2020b}. Interpretation of such an experiment can be simplified if the accurate theoretical prediction of the HFS constants for RaF is available.

Interpretation of the experiments with heavy-atom \textit{molecules} to search for T,P-violating or P-violating effects requires high-precision calculations of a number of molecular constants that are required to express the observed effect in terms of fundamental properties of the nucleus and electron~\cite{Skripnikov:16b,Skripnikov:17c,Fleig:16,Fleig:17,Maison:2019b,Skripnikov:2020b,Flambaum:2020d,Flambaum:2020e,Roussy:2020}. These constants cannot be measured. Therefore, to test the accuracy of their calculation one usually compares the theoretical value of the hyperfine structure constant with the experimental one. All the above characteristics are mainly determined by the behavior of the valence wavefunction of the molecule in the vicinity of the heavy-atom nucleus. However, there was no systematic attempt to take into account the BW contribution in such recent accurate calculations for the molecules of interest. In Ref.~\onlinecite{Malkin:2011} there was an attempt to treat the BW contribution to the hyperfine structure of small molecules within the gaussian distribution model of the nuclear magnetization using the density functional electronic structure calculations.

In the present paper, we develop a method that can be employed to treat the finite nuclear magnetization distribution effect in atoms and \textit{molecules} if the corresponding (nonzero) atomic or molecular data are known for one of the electronic states with sufficient accuracy. This method considers a general model of the  magnetization distribution inside a nucleus, i.e. without using any particular model such as uniformly magnetized ball, gaussian-distributed magnetization model, etc. The proposed method is applied to predict hyperfine structure constants for the $^{225}$RaF molecule and the $^{225}$Ra$^+$ cation in the ground and low-lying excited states within the 4-component relativistic coupled cluster theory. The $^{225}$Ra isotope has the nuclear spin equal to $1/2$ and has only a  magnetic dipole hyperfine structure, i.e. does not exhibit a quadrupole one. This is important from the experimental point of view to unambiguously extract the magnetic hyperfine dipole
interaction~\cite{Garcia:2020b}.

\section{Theory}

For the nucleus with the spin $I$ and the nuclear $g-$factor the nuclear magnetic dipole moment is 
\begin{equation}
\mu=gI\mu_N, 
\end{equation}
where $\mu_N=\frac{e\hbar}{2m_pc}$ is the nuclear magneton, $e$ is the elementary charge, $\hbar$ is the Planck constant and $m_p$ is the proton mass.

To treat the nuclear magnetization distribution inside the finite nucleus one can use the following substitution~\cite{Zherebtsov:2000,Shabaev:1997,Tupitsyn:02}:
\begin{equation}
\mu\to\mu(r)=\mu F(r).
\label{Funct}
\end{equation}
The function $F(r)$ takes into account the nuclear magnetization distribution inside the finite nucleus. In the point magnetic dipole moment approximation $F(r)=1$. The expressions for different models can be found elsewhere  \cite{Zherebtsov:2000,Tupitsyn:02,Malkin:2011}.

The hyperfine interaction of the magnetic moment $\bm{\mu}$ of a given nucleus with electrons is described by the following one-electron operator:
 \begin{equation} 
 \label{hfs1}
 h^{\rm HFS}=\sum_i \frac{\bm{\mu}\cdot[\mathbf{r}_i\times \bm{\alpha}_i]}{r_i^3} F(r_i)\\
 = (\bm{\mu}\,\bm{T}),
 \end{equation}
 where 
 \begin{equation} 
 \label{hfsT}
 \bm{T}=\sum_i \frac{[\mathbf{r}_i\times \bm{\alpha}_i]}{r_i^3} F(r_i),
 \end{equation} 
  $\bm{\alpha}$ are Dirac's matrices:
\begin{eqnarray}
  \bm{\alpha}=
  \left(\begin{array}{cc}
  \bm{0}      & \bm{\sigma} \\
  \bm{\sigma} & \bm{0} \\
  \end{array}\right),
 \label{alph}  
\end{eqnarray}  
  $\bm{\sigma}$ are Pauli matrices and $\mathbf{r}_i$ is the radius-vector of the electron $i$ with respect to the position of the considered nucleus.

For a linear diatomic molecule, one can introduce the HFS constant $A_{||}$ associated with the nucleus $K$ as:
\begin{eqnarray}
\label{Apar}
 A_{||}(K) = \frac{\mu_{K}}{I \Omega}\, 
 \left< \Psi_{\Omega} \right| \sum_i\, \left( \frac{{\bf{r}}_{iK} \times \bm{\alpha}_i}{r_{iK}^3}
 \right)_z F(r_{iK})\left| \Psi_{\Omega} \right>\\
 = \frac{\mu_{K}}{I \Omega}\, 
 \left< \Psi_{\Omega} \right| T_z(K)\left| \Psi_{\Omega} \right>,
\end{eqnarray}
where $\mathbf{r}_{iK}$ is the radius-vector of electron $i$ with respect to the nucleus $K$ and $\Omega$ is the projection of the total electronic angular momentum on the internuclear axis $z$.
 
For a molecule in the electronic state with $|\Omega|=1/2$ one introduces also the $A_{\perp}$ constant:
\begin{eqnarray}
 A_{\perp}(K)  & =  &  \nonumber \\
 \frac{\mu_{K}}{I \Omega}\,
 \left< \Psi_{\Omega=+1/2} \right| \sum_i\, \left( \frac{{\bf{r}}_{iK} \times \bm{\alpha}_i}{r_{iK}^3}
 \right)_+ F(r_{iK})\left| \Psi_{\Omega=-1/2} \right>. & &
 \label{Aperp}
\end{eqnarray}
where the $V_+$ component of some vector $\bf{V}$ means $V_+=V_x + iV_y$. In the present paper we assume that the relative phase of $\Psi_{\Omega=+1/2}$ and $\Psi_{\Omega=-1/2}$ wavefunctions is chosen in such a way that the matrix element of the total electronic angular moment $\left< \Psi_{\Omega=+1/2} \right|  J_+ \left| \Psi_{\Omega=-1/2} \right>$ is positive.

Eq.~(\ref{Apar}) can also be used to define the hyperfine structure constant, $A$, of the atom in the electronic state with the total electronic angular momentum J and its projection $M_J$. For this one should replace $\Omega$ by $M_J$ in Eq.~(\ref{Apar}).

One can use the following expression for the hyperfine structure constant \cite{bohr1950influence, bohr1951bohr},
\begin{equation}
    A= A^{(0)} - A^{\rm BW},
\label{HFSbw}    
\end{equation}
where $A^{(0)}$ is the HFS constant for the point nuclear magnetic moment (i.e. for $F(r)=1$) and $A^{\rm BW}$ is:
\begin{equation}
    A^{\rm BW}=A^{(0)}-A.
\label{Abw}    
\end{equation}
$A^{\rm BW}$ gives contribution of the finite nuclear magnetization distribution to the HFS constant. Note, that for a more direct comparison with the experimental HFS values one should also include a contribution of the QED effects into the hyperfine structure constant, $A^{\rm QED}$. However, such contributions are available only for a limited number of systems. For simplicity, this contribution has been omitted in the above equations but can be added to the  right-hand side of Eq.~(\ref{HFSbw}).

Function $F(r)$ is noticeably different from $1$ only inside the nucleus. Therefore, it follows from Eqs.~(\ref{Apar}) and (\ref{Abw}) that $A^{\rm BW}$ depends on the behaviour of the wavefunction only in this region. One can also introduce the following operator:
 \begin{equation} 
 \label{Tbw}
 \bm{T^{\rm BW}}=\sum_i \frac{[\mathbf{r}_i\times \bm{\alpha}_i]}{r_i^3} (1-F(r_i)).
 \end{equation} 
Thus, 
\begin{eqnarray}
 A^{\rm BW}_{||}(K)  
 = \frac{\mu_{K}}{I \Omega}\, 
 \left< \Psi_{\Omega} \right| T^{\rm BW}_z(K) \left| \Psi_{\Omega} \right>.
 \label{AparBW}
\end{eqnarray}
The operator (\ref{Tbw}) is zero outside the nucleus.

Dirac equation for radial components of the one-particle wavefunction can be written in the following form:
\begin{eqnarray}
\label{DirEq1}  
    \hbar c(\frac{dg_{n\kappa}}{dr}+\frac{1+\kappa}{r}g_{n\kappa})-(E_{n\kappa}+mc^2-V)f_{n\kappa}=0~~~~\\  
\label{DirEq2}       
    \hbar c(\frac{df_{n\kappa}}{dr}+\frac{1-\kappa}{r}f_{n\kappa})+(E_{n\kappa}-mc^2-V)g_{n\kappa}=0,~~~         
\end{eqnarray}
where $g$ and $f$ are radial functions of large and small components of the Dirac bispinor, $\kappa=(-1)^{j+l+1/2}(j+1/2)$, j is the total momentum of the electron, $n$ is the principle quantum number, $V$ is the nuclear potential.
In the nuclear region it is possible to neglect the binding energy $E_{n\kappa}$ with respect to the nuclear potential. Therefore, the wavefunctions with different $n$ and the same $\kappa$ are proportional to each other in this region. This property is widely used in different applications~\cite{Titov:14a,Khriplovich:91,Titov:96,Titov:99, Mosyagin:10a,Mosyagin:16,Shabaev:2006,Dzuba:2011,Skripnikov:16a,Skripnikov:15b}.

From the structure of the HFS Hamiltonian (\ref{hfsT}) one can see that its expectation value for a given bispinor for a system with a spherical symmetry is proportional to $\int_0^{\infty} gf F(r) dr$. At the same time 
$$A^{\rm BW} \propto \int_0^{\infty} gf (1 - F(r)) dr \approx \int_0^{ R_{\rm nuc}} gf (1 - F(r)) dr,$$ where $ R_{\rm nuc}$ is the radius of the nucleus. As only $s_{1/2}$ and $p_{1/2}$ functions have non-negligible amplitudes inside the nucleus only these types of functions can noticeably contribute to~$A^{\rm BW}$.

In the nuclear region $r\sim  R_{\rm nuc}$ of a heavy ion (atom) the absolute value of the nuclear potential $V(r)$ is larger than the electron rest energy and one can omit the $mc^2$ terms in addition to $E_{n\kappa}$ in the equations (\ref{DirEq1}) and (\ref{DirEq2})~\cite{Shabaev:2006}. From the simplified equations it follows that in the region $r\sim  R_{\rm nuc}$ the product $g_{n\kappa}f_{n\kappa}$ differs from $g_{n -\kappa}f_{n -\kappa}$ only by a constant factor~\cite{Shabaev:2006}. For $s_{1/2}$ and $p_{1/2}$ functions it can be also seen from the explicit analytical expressions that are given in Ref.~\onlinecite{Khriplovich:91}. Thus, one obtains the following expression that connects for matrix elements of $z$-component of the operator (\ref{Tbw}), $T^{\rm BW}_z$, over the $1s_{j=1/2,m_j=1/2}$ and $2p_{j=1/2,m_j=1/2}$ functions:
\begin{eqnarray}
\int 2p_{1/2,1/2}^{\dagger} T^{\rm BW}_z 2p_{1/2,1/2} d\mathbf{r}
\approx \nonumber \\
 \beta \int 1s_{1/2,1/2}^{\dagger} T^{\rm BW}_z 1s_{1/2,1/2} d\mathbf{r},
\label{spprop}
\end{eqnarray}
where $\beta$ is the proportionality constant independent on the actual expression for the function $F(r)$. The only suggestion is that $(F(r)-1)$ is localised inside the nucleus. Similar idea has been used in Ref.~\onlinecite{Shabaev:2006} to introduce the specific difference of the electronic $g$-factors of hydrogen-like and boron-like ions of lead.

Let us now consider a reduced one-particle density matrix $\rho(\mathbf{r}|\mathbf{r'})$ obtained from correlation calculation of some molecule containing a heavy nucleus K and choose the origin at the position of K. One can write:
\begin{equation}
\begin{array}{l}                 
\rho(\mathbf{r}|\mathbf{r'}) = \sum\limits_{p,q} \rho_{p,q} \varphi_{p}(\mathbf{r})\varphi_{q}^{\dagger}(\mathbf{r'}),
 \label{rrofull}
\end{array}
\end{equation}
where $\{\varphi_{p}\}$ are molecular bispinors.
The expectation value of some one-particle operator $X$ can be calculated as:
\begin{equation} 
   \langle {X} \rangle = \sum_{p,q} \rho_{p, q}\int \varphi_{q}^{\dagger}X\varphi_{p}d\mathbf{r}\ .
 \label{OperFull} 
\end{equation} 
In this paper we are interested in the operator $\mathbf{T^{\rm BW}}$ (\ref{Tbw}), which is zero outside the nucleus. It means that for such operator $X$ one can perform integration in Eq.~(\ref{OperFull}) inside the sphere of radius $R_c \approx  R_{\rm nuc}$. In this region one can reexpand  $\{\varphi_{p}\}$ in terms of some sufficiently complete set of basis functions $\eta_{nljm}(\mathbf{r})$ centered at the nucleus K:
\begin{equation} 
   \varphi_{p}(\mathbf{r}) \approx \sum_{nljm} C^p_{nljm} \eta_{nljm}(\mathbf{r}),~~|\mathbf{r}| \le R_c,
 \label{expans}     
\end{equation} 
where $C^p_{nljm}$ are expansion coefficients. For example, one can take hydrogen-like functions for the ion with the nucleus K as a set of $\{\eta_{nljm}\}$. From the above consideration, such functions with a given set of $l,j,m$ and different $n$ are proportional to each other inside the nucleus.
Therefore, for each combination of $l,j,m$ one can introduce some \textit{reference function} $\eta_{ljm}(\mathbf{r})$ and write:
\begin{eqnarray}
    \eta_{nljm}(\mathbf{r}) \approx k_{nljm} \eta_{ljm}(\mathbf{r})\ ,\ \  |\mathbf{r}|\le R_c\ ,
 \label{proport}     
\end{eqnarray}
where $k_{nljm}$ is the proportionality coefficient. 
Functions $\eta_{ljm}(\mathbf{r})$ with different $m$ and the same $l,j$ differ only in their spin-angular part. Substituting Eqs.~(\ref{expans}) and (\ref{proport}) into Eq.~(\ref{OperFull}) and taking into account that the integration in Eq.~(\ref{OperFull}) can be performed inside the sphere of radius $R_c$ with center at the nucleus K one obtains:
\begin{equation}
 \langle {X} \rangle\  \approx \sum\limits_{ljm;l'j'm'} {\cal{P}}_{ljm, l'j'm'} \int\limits_{|\mathbf{r}|<R_c} \eta_{l'j'm'}^{\dagger}X\eta_{ljm} d\mathbf{r}, 
 \label{OperRedFinFin} 
\end{equation} 
where
\begin{equation} 
\begin{array}{l} 
{\cal{P}}_{ljm, l'j'm'}=
\sum\limits_{p,q,n,n'} \rho_{p,q} C^{p}_{nljm}k_{nljm} C^{q*}_{n'l'j'm'}k^{*}_{n'l'j'm'}.
 \label{WRed} 
\end{array}  
\end{equation}
${\cal{P}}_{ljm, ljm}$ can be called the reduced occupation associated with the reference function $\eta_{ljm}$.
Such parameters are introduced in the atoms in compounds theory~\cite{Titov:14a} (see also Refs.~\cite{Lomachuk:13,Zaitsevskii:16a,Oleynichenko:2018,Skripnikov:15b}) as they can serve as some certain characteristics of an atom inside a compound.

As it was noted above, inside the nucleus only $s_{1/2}$ and $p_{1/2}$ functions have non-negligible amplitudes and can contribute to $A^{\rm BW}$. Therefore, in the present case one can take $1s_{1/2}$ and $2p_{1/2}$ functions of the hydrogen-like ion with the nucleus $K$ as reference functions ${\eta_{ljm}(\mathbf{r})}$.
One can also take into account the following relation for $z$-component of the operator (\ref{Tbw}) (and (\ref{hfsT})):
\begin{equation}
\int \eta_{ljm}^{\dagger}T^{\rm BW}_z\eta_{ljm} d\mathbf{r}=-\int \eta_{lj-m}^{\dagger}T^{\rm BW}_z\eta_{lj-m} d\mathbf{r}.
\end{equation}
Besides, off-diagonal matrix elements of the z-component of operators (\ref{hfsT}) and (\ref{Tbw}) between $s_{1/2}$ and $p_{1/2}$ functions are zero.
Introducing also
\begin{eqnarray} 
\label{Ps}
{\cal{P}}_s={\cal{P}}_{1s_{1/2,1/2},~1s_{1/2,1/2}} - {\cal{P}}_{1s_{1/2,-1/2},~1s_{1/2,-1/2}},\\
\label{Pp}
{\cal{P}}_p={\cal{P}}_{2p_{1/2,1/2},~2p_{1/2,1/2}} - {\cal{P}}_{2p_{1/2,-1/2},~2p_{1/2,-1/2}},\\
\label{Bs}
B_s=\int\limits_{|\mathbf{r}|\le R_{\rm nuc}} \eta_{1s_{1/2,1/2}}^{\dagger}T^{\rm BW}_z\eta_{1s_{1/2,1/2}} d\mathbf{r}, \\
B_p = \int\limits_{|\mathbf{r}|\le R_{\rm nuc}} \eta_{2p_{1/2,1/2}}^{\dagger}T^{\rm BW}_z\eta_{2p_{1/2,1/2}} d\mathbf{r},
\end{eqnarray} 
one obtains:
\begin{eqnarray} 
\langle {T^{\rm BW}_z} \rangle\ {\approx}~ 
{\cal{P}}_s \int\limits_{|\mathbf{r}|\le R_{\rm nuc}} \eta_{1s_{1/2,1/2}}^{\dagger}T^{\rm BW}_z\eta_{1s_{1/2,1/2}} d\mathbf{r}~~ \nonumber
\\     
      + ~     {\cal{P}}_p \int\limits_{|\mathbf{r}|\le R_{\rm nuc}} \eta_{2p_{1/2,1/2}}^{\dagger}T^{\rm BW}_z\eta_{2p_{1/2,1/2}} d\mathbf{r}~~\\
=
{\cal{P}}_s B_s + {\cal{P}}_p B_p.~~ \nonumber
 \label{Op1} 
\end{eqnarray} 
Now we can rewrite Eq.~(\ref{AparBW}) as:
\begin{eqnarray}
 A^{\rm BW}_{||}(K)  
 \approx \frac{\mu_{K}}{I \Omega} ({\cal{P}}_s B_s + {\cal{P}}_p B_p)
 \label{AparBW2}.
\end{eqnarray}
Using Eq.~(\ref{spprop}) it is possible to simplify Eq.~(\ref{AparBW2}):
\begin{eqnarray}
 \label{AparBW3}
 A^{\rm BW}_{||}(K)  
 \approx \frac{\mu_{K}}{I \Omega} ({\cal{P}}_s + \beta{\cal{P}}_p)B_s \\
  = A^{\rm BW,s}_{||}(K) + A^{\rm BW,p}_{||}(K), 
\end{eqnarray}
where 
\begin{eqnarray}
\label{AparBWs}
 A^{\rm BW,s}_{||}(K)  
 = \frac{\mu_{K}}{I \Omega} {\cal{P}}_s B_s, \\
\label{AparBWp} 
 A^{\rm BW,p}_{||}(K)  
 = \frac{\mu_{K}}{I \Omega} {\cal{P}}_p \beta B_s.
\end{eqnarray}
Corresponding expression can be written for $A^{\rm BW}_{\perp}(K)$ for the diatomic molecule in the $\Omega=1/2$ electronic state as well as $A^{\rm BW}$ in case of a single atom. Generalizations are also possible for polyatomic molecules.

In some cases it can be more convenient to introduce functions $h_{ljm}(\mathbf{r})$, defined as:
\begin{equation}
  h_{ljm}(\mathbf{r})=\eta_{ljm}(\mathbf{r}) \theta(R_c-|\mathbf{r}|),
\end{equation}
where $\theta(R_c-|\mathbf{r}|)$ is the Heaviside step function:
\[
\theta(R_c-|\mathbf{r}|) = \left\{\begin{array}{l}
1,\  |\mathbf{r}|<R_c\\
0,\  \mbox{otherwise.}
\end{array}\right.
\]
Using functions $h(\mathbf{r})$ it is possible to calculate ${\cal{P}}_{ljm, l'j'm'}$ as a mean value of the following operator:
\begin{equation}
    R_{ljm, l'j'm'} = \frac{|h_{ljm}><h_{l'j'm'}|}{<h_{ljm}|h_{ljm}><h_{l'j'm'}|h_{l'j'm'}>}.
\label{Proj}    
\end{equation}

Constants ${\cal{P}}_s$ and ${\cal{P}}_p$ in Eq.~(\ref{AparBW3}) are determined by the electronic structure of a heavy atom or a molecule under consideration and do not depend on the nuclear magnetization distribution model. In contrast, the $B_s$ constant is common for a given heavy atom, its ion or a molecule, containing this atom. It directly depends on the magnetization distribution inside the nucleus under consideration. Therefore, if one has sufficiently accurate theoretical data for the electronic part of the problem of calculation HFS constant for a given heavy atom, ion or a molecule containing this atom as well as accurate experimental data, it is possible to extract the $B_s$ constant value. It can be further used to predict the finite nuclear magnetization distribution effect associated with a given nucleus in any other system that contains this nucleus, i.e. an atom, ion or molecule in any other electronic state. It can also be used to test predictions of the nuclear models using sufficiently accurate atomic data.

Note, that the suggested approach can be used to approximately calculate the specific difference parameter $\xi$ introduced in Ref.~\onlinecite{Shabaev:01a} for lithium-like and hydrogen-like ions using Eq.~(\ref{AparBW3}).

In the present paper we have implemented theoretical approach to calculate ${\cal{P}}_s$ and ${\cal{P}}_p$ constants in Eq.~(\ref{AparBW3}).

\section{Computational details}
All calculations have been performed within the Dirac-Coulomb Hamiltonian. To solve electronic many-body problems for the atom and molecule under consideration we have used the coupled cluster with single, double, and perturbative triple cluster amplitudes method, CCSD(T)~\cite{Stanton:97}. The energy cutoff for virtual
orbitals was set to 10000 Hartree in the correlation treatment. In Ref.~\onlinecite{Skripnikov:17a} it was demonstrated that such an energy cutoff is important to ensure including functions that describe spin-polarization and correlation effects for inner-core electrons. To treat higher-order correlation effects for valence and outer-core electrons we have employed the coupled cluster method with inclusion up to triple and perturbative quadruple cluster amplitudes, CCSDT(Q)~\cite{Kallay:1}. 

To be able to include the most important basis functions with high $lj$ in the 4-component correlation calculation we have used the method of constructing natural basis sets~\cite{Skripnikov:13a}. This method implies scalar-relativistic correlation calculation using a large basis set followed by a diagonalization of atomic blocks of one-particle density matrix~\cite{Skripnikov:13a}. In the present paper, this method has been extended in the following way. For a given $l(j)$ several natural basis functions with the largest occupations have been chosen. These functions are contracted, i.e. expanded in terms of a large number of primitive gaussian-type functions. These contracted functions have been reexpanded in terms of a small number of primitive gaussian-type functions. In the final calculation, we include these primitive gaussians instead of original natural contracted basis functions. Such an approach allows one to have the additional flexibility of the basis set as well as use it in the 4-component calculations. 

For atomic calculations the following uncontracted gaussian-type basis sets have been used. The LBas basis set consisting of 38 $s$--, 33 $p$--, 24 $d$--, 14 $f$--, 7 $g$--, 3 $h$-- and 2 $i$-- functions, which can be written as (38s,33p,24d,14f,7g,3h,2i). This basis set has been obtained by augmentation of the uncontracted Dyall's AE3Z basis set~\cite{Dyall:12} by 5 $s$, 4 $p$, 5 $d$ and 1 $f$ diffuse functions.
Also we have added 3 $g$, 2 $h$ and 2 $i$ functions generated using the method of constructing natural basis sets described above. To treat high-order correlation effects, we have used the SBas basis set that is equal to the uncontracted Dyall's VDZ~\cite{Dyall:12} basis set augmented by a few $s$--, $p$-- and $d$-- type function. To test basis set completeness we have also employed the  
LBasExt basis set: (42s,38p,27d,17f,11g,3h,2i). This basis set has been obtained by augmentation of the uncontracted Dyall's AE4Z basis set~\cite{Dyall:12} by 5 $s$, 4 $p$, 4 $d$ and 1 $f$ diffuse functions and 3 $g$, 2 $h$ and 2 $i$ natural functions.

For calculation of HFS constants for the RaF molecule, we have used the LBasRaF basis set that corresponds to the LBas basis set on Ra and the uncontracted Dyall's AETZ~\cite{Dyall:12} basis set on F. The LBasExtRaF basis set corresponds to the LBasExt basis set for Ra and 
Dyall's AE4Z~\cite{Dyall:12} basis set on F. Finally, we have used the SBasRaF basis set that corresponds to the SBas basis set for Ra
and the aug-cc-PVDZ-DK basis set \cite{Dunning:89,Kendall:92} for~F.

In calculations of the ground and first excited electronic states of RaF we have used the calculated values of equilibrium internuclear distances: R(Ra--F)=4.23 Bohr for both states, obtained in this work. This distance is in agreement with the previous study~\cite{Kudashov:14}.

In the electronic structure calculations, the Gaussian charge distribution model \cite{visscher1997dirac} has been employed. For molecular calculations the {\sc dirac}~\cite{DIRAC15} and {\sc mrcc}~\cite{MRCC2020,Kallay:1} codes have been used. To calculate ${\cal{P}}_s$ and ${\cal{P}}_p$ parameters the code developed in this work has been employed.

The $^{225}$Ra nucleus has spin $I=1/2$ and magnetic moment $\mu$=-0.7338(15) obtained in the direct (atomic) measurement~\cite{Arnold:1987}.

\section{Results and discussion}

There is a strong uncertainty in calculating the BW correction for heavy atoms from first principles due to 
a rather limited knowledge of the nuclear magnetization distribution $F(r)$.
However, as it can be seen from Eq.~(\ref{AparBW3}) the BW contribution to the hyperfine structure constant of an atom or a \textit{molecule} depends only on one matrix element $B_s$ given by Eq.~(\ref{Bs}). The latter element depends on $F(r)$ through Eq.~(\ref{Tbw}).
Thus, if there are accurate experimental data for the (nonzero) hyperfine structure constant for some system induced by a given nucleus as well as accurate enough electronic structure calculation of the HFS constant in the point magnetic dipole approximation it is possible to extract the $B_s$ constant and use it for further predictions. Below we perform this extraction from the atomic data for the $^{225}$Ra$^+$ cation to accurately predict hyperfine structure constants for the $^{225}$RaF molecule.

\subsection{$^{225}$Ra$^+$ cation}
The hyperfine structure constant for the ground 7s~$^2$S$_{1/2}$  state of the $^{225}$Ra$^+$ cation is known with high accuracy~\cite{Neu:1988,Wendt:1986,Dammalapati:2016}. Moreover, for this system, even the QED contribution to the hyperfine structure constant has been recently calculated in Ref.~\onlinecite{ginges2017ground}. Table~\ref{TResRa} gives the calculated values of the hyperfine structure constant $A^{(0)}$ for the ground and excited electronic states of the $^{225}$Ra$^+$ cation calculated in the point magnetic dipole approximation, i.e. without the BW correction. Calculated values are in good agreement with previous theoretical works~\cite{ginges2017ground,Dzuba:1985,Heully:1985,Andriessen:1992,Xing:1995,Sudip:2016b}.

\begin{table}[]
\caption{Hyperfine structure constant (in MHz) for the ground and excited states of $^{225}$Ra$^+$ cation calculated in the point magnetic dipole approximation.}
\begin{tabular}{lrrr}
\hline
\hline
Method                                   & 7s $^2$S$_{1/2}$ & ~~7p $^2$P$_{1/2}$ & ~~7p $^2$P$_{3/2}$ \\ 
\hline
DHF                                      & -21976 & -3657 & -276  \\
CCSD                                     & -29160 & -5484 & -459  \\
\\
CCSD(T)                                  & -28896 & -5498 & -463  \\
correlation correction                   & -117   & -31   & 0    \\
Basis set correction                     &   2   &  3   & 0    \\
Total, electronic ($A^{(0)}$)            & -29012 & -5526 & -463  \\
\hline
\hline
\end{tabular}
\label{TResRa}

\end{table}

The leading contributions to the HFS constants given in Table~\ref{TResRa} have been calculated within the 4-component CCSD(T) approach using the LBas basis set. All 87 electrons of the Ra$^+$ ion have been included in the correlation treatment. For comparison, Table~\ref{TResRa} gives also the values obtained within the Dirac-Hartree-Fock (DHF) and CCSD approaches. Higher order correlation effects have been estimated as a difference of the CCSDT(Q) and CCSD(T) values obtained within the SBas basis set and correlating 27 outer electrons of Ra$^+$. The basis set correction given in  Table~\ref{TResRa} is the contribution of the basis set extension up to the LBasExt basis set. It has been calculated within the 59-electron CCSD(T) approach, i.e. $1s..3d$ electrons have been excluded from the correlation treatment.

The final value of the HFS constant calculated in the point magnetic dipole approximation ($A^{(0)}$) for the ground state of $^{225}$Ra$^+$ given in Table~\ref{TResRa} is in excellent agreement (within 0.03\%) with the most recent previous theoretical value for this constant from Ref.~\onlinecite{ginges2017ground}. It can be seen that the basis set corrections for the ground and excited states HFS constants are small. This confirms the quality of the main LBas basis set. From the values given in Table~\ref{TResRa}, one can expect that the theoretical uncertainty of the HFS constant, calculated in the point magnetic dipole approximation for the ground state of $^{225}$Ra$^+$,
$A^{(0)}$ (see Eq.~(\ref{HFSbw})), is about 150 MHz or less than 1\%. Finite nuclear magnetization distribution contribution has not been included in the uncertainty as it is related to the constant $A$ (see below). Breit and QED effects, calculated in Ref.~\onlinecite{ginges2017ground} contribute in total about 66 MHz. Taking this into account and using the calculated value of $A^{(0)}$ from Table~\ref{TResRa} and experimental value~\cite{Neu:1988} of $A$ 
one finds: 
$$A^{\rm BW}(^{225}{\rm Ra}^+, 7s~^2S_{1/2})=-1215~{\rm MHz}.$$
This is about an order of magnitude bigger than the QED and Breit effects and more than 4\% of the total HFS constant. 

In Refs.~\cite{ginges2017ground,Dzuba:1985,Heully:1985,Panigrahy:1991} there were attempts to obtain $A^{\rm BW}$ within the concrete nuclear magnetization distribution models for different isotopes of Ra. In the present work we have not used any nucleus model to calculate A$^{\rm BW}$ and extracted it from the accurate experimental and theoretical data for the ground state of $^{225}$Ra$^+$.

\begin{table}[]
\caption{BW contributions $A^{BW}, A^{\rm BW,s}, A^{\rm BW,p}$ and the final values of the hyperfine structure constants (in MHz) for the ground and excited states of the $^{225}$Ra$^+$ cation. For the ground state $A^{\rm BW}$ has been obtained as a difference between the theoretical value of the HFS constant calculated in the point magnetic dipole approximation and the experimental value taking into account QED and Breit effects.}
\begin{tabular}{lrrr}
\hline
\hline
                                         & 7s $^2$S$_{1/2}$ & ~~7p $^2$P$_{1/2}$ & ~~7p $^2$P$_{3/2}$ \\ 
\hline
$-A^{\rm BW,s}$                              & 1214     & -5    & 3       \\
$-A^{\rm BW,p}$                              & 1        & 80    & 0       \\
$-A^{\rm BW}$                                & 1215   & 75    & 2       \\
                                         &          &       &           \\
$A^{(0)}$ (see Table~\ref{TResRa})       & -29012 & -5526 & -463  \\
Breit+QED*, Ref.~\onlinecite{ginges2017ground} & 66(23)   & ---   & ---       \\
                                         &          &       &           \\
Final                                    & -27731   & -5451 & -461   \\
Experiment~\cite{Neu:1988,Wendt:1986,Dammalapati:2016}       & -27731(13) & -5446.0(7) & -466.4(4.6)   \\
\hline
\hline
\end{tabular}
\label{TResRaBW}

*Extracted from Ref.~\onlinecite{ginges2017ground}: Breit: -93 MHz; QED: 159(23) MHz; Electron+Breit:-29113 MHz.\\

\end{table}

According to calculations, $1s..3d$ electrons of Ra contribute about 2.3\% to the HFS constant $A^{(0)}$ of the ground state of $^{225}$Ra$^+$. $1s..4f$ electrons of Ra contribute about 4\%. In Ref.~\onlinecite{Skripnikov:17a} a similar correlation contribution 
to the enhancement factor of the T,P-violating scalar-pseudoscalar nucleus-electron interaction has been found for the Fr atom. 
It is interesting to note that while the inclusion of $1s..4f$ electrons in the correlation treatment increases the absolute value of the HFS constant, the BW effect in the ground electronic state of $^{225}$Ra decreases it by 4\%. 
Therefore, it is possible to obtain an ``excellent agreement'' (e.g. less than 1\%) of the calculated HFS constant $A^{(0)}$ in the simplified calculation (with the exclusion of BW contribution and correlation contribution of the core electrons) with the experimental HFS constant~$A$. But this will not mean that the atomic enhancement factor of the T,P-violating effect is calculated with such accuracy (less than 1\%).

As it was noted above, $B_s$ (as well as $\beta$) constant does not depend on the electronic state of the system containing heavy nucleus under consideration. Therefore, it is possible to use Eq.~(\ref{AparBW3}) to express the BW contribution to the hyperfine structure constant of the system containing the $^{225}$Ra nucleus 
in terms of $A^{\rm BW}(^{225}{\rm Ra}^+, 7s~^2S_{1/2})$ by calculating parameters ${\cal{P}}_s$ and  ${\cal{P}}_p$ for the $^{225}{\rm Ra}^+$ in $7s~^2S_{1/2}$ state as well as for the system under consideration. Such calculations have been performed for the excited states of the $^{225}{\rm Ra}^+$. It was found (see Table~\ref{TResRaBW}):
$$A^{\rm BW}(^{225}{\rm Ra}^+, {\rm 7p}~^2P_{1/2})=-75~{\rm MHz}.$$
Using Eq.~(\ref{HFSbw}) we obtain 
$$A(^{225}{\rm Ra}^+, {\rm 7p}~^2P_{1/2})=-5451~{\rm MHz}.$$
This value is in a far better agreement with the experimental value than the value of $A^{(0)}(^{225}{\rm Ra}^+, {\rm 7p}~^2P_{1/2})$ given in Table~\ref{TResRa}. Note, that there can be some contribution from the QED (and Breit) effects. No such contributions have been calculated for the 7p $^2$P$_{1/2}$ state of Ra$^+$, but one can expect that they are small taking into account the corresponding value for the ground state of Ra$^+$ (0.2\%~\cite{ginges2017ground}). Table~\ref{TResRaBW} gives the calculated values of $A^{\rm BW}$ for all of the considered states of $^{225}$Ra$^+$. Table~\ref{TResRaBW} also gives contributions $A^{\rm BW,s}$ and $A^{\rm BW,p}$ due to $s_{1/2}$ and $p_{1/2}$ terms (see Eq.~(\ref{AparBW3})).
It can be seen that for the 7s $^2S_{1/2}$ state the former contribution dominates, while for 7p $^2P_{1/2}$ the latter contribution dominates. The BW contribution to HFS constant of the 7p $^2P_{3/2}$ state is very small in contrast to the case of the 6p $^2P_{3/2}$ state of the Tl atom~\cite{Prosnyak:2020}.

In the above treatment of the $A^{\rm BW}$ constant no specific nuclear distribution model has been used. However, if we assume some nuclear magnetization distribution model by specifying $F(r)$ in Eq.~(\ref{Funct}) then the $A^{\rm BW}$ constant can be calculated either (i) directly using Eq.~(\ref{AparBW}) or (ii) in the way suggested by Eq.~(\ref{AparBW3}). In the latter case the ``nuclear part'' reduces to just one matrix element $B_s$ given by Eq.~(\ref{Bs}) on $1s_{1/2}$ function of the corresponding hydrogen-like ion. In our case one can consider matrix element for the ground 1s-state of the hydrogen-like $^{225}$Ra$^{87+}$ ion. Note, that the approach (ii) should be valid not only for the 7s $^2$S$_{1/2}$ state of Ra$^+$ but also for the 7p~$^2$P$_{1/2}$ state and other states. To check the accuracy of the approach (ii) numerically we have considered the uniformly magnetized ball model of radius $R_m$. In this model $F(r)=R_m^3/r^3$ for $r \le R_m$ and $F(r)=1$ for $r > R_m$. Here $R_m \sim R_{\rm nuc}$ can be considered as a constant parameter of the model~\cite{Prosnyak:2020}. Within this model we have found that the values of the $A^{\rm BW}$ constant calculated using the direct approach (i) and using the 
approach (ii) coincide within 0.4\% for all of the considered states of $^{225}$Ra$^+$, i.e. 7s $^2$S$_{1/2}$, 7p $^2$P$_{1/2}$ and 7p $^2$P$_{3/2}$. As it was noted above $A^{\rm BW}$(7p $^2$P$_{1/2}$) is determined by the second term in Eq.~(\ref{AparBW3}) that uses property (\ref{spprop}). This numerical test verifies the suggested approach (ii), i.e. the use of Eq.~(\ref{AparBW3}). Thus, to calculate BW contributions one needs to know the matrix element $B_s$ given by Eq.~(\ref{Bs}) and perform corresponding electronic structure calculations of ${\cal{P}}_s$ and ${\cal{P}}_p$ given by Eqs.~(\ref{Ps}) and (\ref{Pp}).

\subsection{$^{225}$RaF molecule}
The ground $X~^2\Sigma_{1/2}$ as well as the first excited $A~^2\Pi_{1/2}$ electronic states of the RaF molecule qualitatively have one unpaired electron on the so-called non-bonding (atomic-like) orbital. 
This results in an almost diagonal Franck-Condon matrix element between these states. Due to this feature and the fact that the transition frequency between these states lies in the visible region, it is possible to laser cool this molecule~\cite{Isaev:2010,Ruiz:2019}.

From the above consideration of the $^{225}$Ra$^+$ cation, it is expected that the non-negligible contribution of the finite nuclear magnetization distribution to the hyperfine structure constants can be expected also for $^{225}$RaF. This effect has not been accurately considered for this molecule before. Note however, that the need to consider this effect arises only when the calculation of the HFS constant in the point magnetic dipole approximation is accurate enough, i.e. its uncertainty is smaller than the effect under consideration.

Table~\ref{TResRaF} gives the calculated values of the hyperfine structure constant for the ground $X ^2\Sigma_{1/2}$ and first excited $A ^2\Pi_{1/2}$ electronic states of the $^{225}$RaF molecule. The main calculation of $A^{(0)}_{||}$  has been performed within the 4-component CCSD(T) approach using the LBasRaF basis set and correlating all 97 electrons. To calculate $A^{(0)}_{\perp}$, $1s..3d$ electrons of Ra have been excluded from correlation calculation. However, the correcting scaling factor for the treatment of these electrons (obtained from the $A_{||}$ calculation) has been applied
~\cite{Note1}.
Higher-order correlation effects have been estimated as a difference of the CCSDT and CCSD(T) values obtained within the SBasRaF basis set and correlating 27 outer electrons of RaF. We have also calculated the vibrational correction to the considered parameters for the case of zero vibrational levels of the considered electronic states of the $^{225}$RaF molecule using the approach described in Ref.~\cite{Skripnikov:15a}. Finally, we have applied the basis set correction given in Table~\ref{TResRaF}. It is the contribution of the basis set extension up to the~LBasExtRaF basis set. It has been calculated within the 69-electron CCSD(T) approach, i.e. $1s..3d$ electrons of Ra have been excluded from the correlation calculation.

One can expect that the uncertainties of the calculated HFS constants $A^{(0)}_{||}$ and $A^{(0)}_{\perp}$ are of order 1\%  similar to the atomic case.

\begin{table}[!h]
\caption{Hyperfine structure constants $A_{||}$ and $A_{\perp}$ (in MHz) for the ground $X ^2\Sigma_{1/2}$ and excited $A ^2\Pi_{1/2}$ states of the $^{225}$RaF molecule induced by the $^{225}$Ra nucleus.}
\begin{tabular}{lrrrr}
\hline
\hline
                            & \multicolumn{2}{c}{$X ^2\Sigma_{1/2}$}  &  \multicolumn{2}{c}{~~~$A ^2\Pi_{1/2}$}  \\ 
Method                      & $A_{||}$  &  $A_{\perp}$ & $A_{||}$  &  $A_{\perp}$  \\ 
\hline     
DHF                         & -12048    &  ~~~-11670 & ~~~~~~-1638  & -1235  \\               
CCSD                        & -17814    &  -17148    & -2848        & -2173  \\
                            &           &            &              &       \\
CCSD(T)                     & -17595    &  -16941    & -2842        & -2198  \\
correlation correction      & -134      &  -134     & -48          & -28   \\
basis set correction        &  -31      &  -30*      &  -4           &  -3*   \\
vibrational correction      & -19       &  -18*      & -2           &  -2*   \\
Total, electronic ($A^{(0)}_{||/\perp}$ )  & -17780    &   -17123 & -2896  & -2230 \\
                            &           &            &              &       \\
$-A^{\rm BW,s}_{||}$        &  723      &            &    9         &       \\
$-A^{\rm BW,p}_{||}$        &   7       &            &    34        &       \\
$-A^{\rm BW}_{||~/~\perp}$  &  730      &    720*    &    44        & 26*   \\
\\
Final                       & -17049    &   -16403   &   -2852      & -2204  \\
\hline
\hline
\multicolumn{5}{l}{%
    *Rescaled from $A_{||}$.
}\\
\end{tabular}
\label{TResRaF}
\end{table}

The values of the ground state hyperfine constants calculated in the point magnetic dipole approximation are in good agreement with previous calculations~\cite{Kudashov:14,Sudip:2016b}, but
have much smaller uncertainties
~\cite{Note2}.

To calculate $A^{\rm BW}_{||}$ we have calculated ${\cal{P}}_s$ and ${\cal{P}}_p$ constants for the $^{225}$RaF molecule. The resulting value of $A^{\rm BW}_{||}$ as well as its contributions $A^{\rm BW,s}_{||}$ and $A^{\rm BW,p}_{||}$ are given in Table~\ref{TResRaF}. It can be seen that the finite nuclear magnetization distribution effect $A^{\rm BW}_{||}$ contributes about 4\% to the hyperfine structure of the ground state of the $^{225}$RaF molecule and cannot be neglected in an accurate consideration. The leading contribution to $A^{\rm BW}_{||}$ comes from $A^{\rm BW, s}_{||}$. A similar effect and its contributions have been found for the ground state of the $^{225}$Ra$^+$ cation above.

For the first excited state $A ^2\Pi_{1/2}$, $A^{\rm BW}_{||}$ contributes about 1.5\%. Here $A^{\rm BW, p}_{||}$ dominates. This is close to what was found for the excited 7p $^2$P$_{1/2}$ state of $^{225}$Ra$^+$. This also correlates with the assumed atomic-like character of the ground and the first excited states of RaF. Note that the $A^{\rm BW, p}_{||}$ domination over $A^{\rm BW, s}_{||}$ is not so strong as in the atomic case of the 7p $^2$P$_{1/2}$ state. The relative asymmetry in HFS constants $A_{||}$ and $A_{\perp}$ is larger for the excited state 7p $^2$P$_{1/2}$ than for the ground state. It indicates some deviation from the idealized atomic character of this state.

For the ground state of the $^{225}$RaF molecule we have found that the correlation contribution of $1s..4f$ electrons of Ra and $1s$ of F is about 4\% which is close by the absolute value but has an opposite sign to the finite nuclear magnetization distribution effect. Thus, the situation is similar to what was found for the case of $^{225}$Ra$^+$ ground state HFS above. It is possible to obtain an excellent agreement of the calculated HFS constant $A_{||}^{(0)}$ in the simplified calculation with the exclusion of finite nuclear magnetization distribution contribution and electron correlation contribution of the core electrons with the experimental HFS constant~$A_{||}$. But this will not mean that the effective electric field acting on the electron electric dipole moment and other similar parameters of the T,P-violating effects are calculated with such good accuracy using the same wavefunction. It should be stressed also, that the proper correlation of the core electrons requires to use sufficiently high energy cutoff, see Fig. I in Ref.~\onlinecite{Skripnikov:17a} and Ref.~\onlinecite{Skripnikov:15a}.

\section{Conclusions}
The theoretical method to treat the contribution of the finite nuclear magnetization distribution to the hyperfine structure constants of heavy atoms and molecules containing heavy atoms has been proposed. It is shown that the nuclear part of the problem can be reduced to the treatment of just one matrix element over the hydrogen-like function. This matrix element can be calculated using some nuclear structure model or it can be extracted by combining accurate experimental and theoretical data for an atom, (highly charged) ion, or a molecule containing the heavy nucleus under consideration.

An important feature of the formulated theory is that it is possible to separate the electron correlation problem and the nuclear problem using Eq.~(\ref{AparBW3}).  The accuracy of such separation (factorization) has been tested numerically for one of the nuclear magnetization models and found to be very high, the corresponding theoretical uncertainty is less than 1\%.

Most of the previous studies of the effects of the finite distribution of nuclear magnetization concerned atoms. The approach developed here helps to extend this treatment to molecular physics.

The method has been applied to study the hyperfine structure for the $^{225}$Ra$^+$ ion and $^{225}$RaF molecule in low-lying electronic states. In particular, it was found that the finite nuclear magnetization distribution effect strongly (more than 4\%) contributes to the HFS of the ground state of $^{225}$RaF and should be taken into account in precise calculations. This effect has an opposite sign to the correlation contribution of the core electrons, but similar absolute value. Therefore, the neglect of both these effects can lead to an incorrect conclusion about the accuracy of the calculated wavefunction and uncertainties of other properties calculated with this wavefunction. This conclusion should be taken into account in most precise calculations, e.g. in the field of the search of the P- or/and T,P-violation effects in molecules, etc.

Predicted HFS constants for $^{225}$RaF can be used to simplify the interpretation of further experiments with this molecule~\cite{Garcia:2020b}.

\begin{acknowledgments}
I am grateful to A.V. Titov, V.M. Shabaev, M.G. Kozlov and T.A. Isaev for useful discussions. 
Electronic structure calculations have been carried out using computing resources of the federal collective usage center Complex for Simulation and Data Processing for Mega-science Facilities at NRC “Kurchatov Institute”, http://ckp.nrcki.ru/, and computers of Quantum Chemistry Lab at NRC ``Kurchatov Institute" - PNPI.

The data that support the findings of this study are available from the corresponding author upon reasonable request.

Research has been supported by the Russian Science Foundation Grant No. 19-72-10019.

\end{acknowledgments}


\end{document}